\documentstyle[NATO,numreferences]{crckapb}

\begin{opening}
\title{QUANTUM SYSTEMS IN WEAK GRAVITATIONAL FIELDS}

\author{G. PAPINI}
\institute{Department of Physics, University of Regina \\Regina, Sask. S4S 0A2, Canada}
\institute{International Institute for Advanced Scientific Studies \\Vietri sul Mare
(SA), Italy} \institute{Canadian Institute for Theoretical Astrophysics \\University of
Toronto, On., Canada}

\end{opening}

\runningtitle{QUANTUM SYSTEMS IN WEAK GRAVITATIONAL FIELDS}

\begin{document}


\section{Introduction}

Fully covariant wave equations predict the existence of a  class of
inertial-gravitational effects that can be tested experimentally. In these equations
inertia and gravity appear as external classical fields, but, by conforming to general
relativity, provide very valuable information on how Einstein's views carry through in
the world of the quantum. Experiments already confirm that inertia and Newtonian gravity
affect quantum particles in ways that are fully consistent with general relativity down
to distances of $\sim 10^{-4}cm$ for superconducting electrons \cite{hildebrandt} and of
$\sim 10^{-8}cm$ for neutrons \cite{colella,werner,bonse}. Other aspects of the
interaction of gravity with quantum systems are just beginning to be investigated.

Gravitational-inertial fields in the laboratory are weak and remain so in the cosmos for
most astrophysical sources. These are the fields considered here. They are adequately
described by the weak field approximation (WFA).

  Gravitational-inertial fields affect particle wave functions in a
variety of ways. They induce quantum phases that afford a unified treatment of
interferometry and gyroscopy. They interact with particle spins giving rise to a number
of significant effects. They finally shift energy levels in particle spectra
\cite{parker}. While it still is difficult to predict when direct measurements will
become possible in the latter case, rapid experimental advances in particle
interferometry \cite{opat,kaiser,borde} require that quantum phases be derived with
precision. This will be done below for Schroedinger, Klein-Gordon, Maxwell and Dirac
equations. Large, sensitive interferometers hold great promise in many of these
investigations. They can play a role in testing general relativity.

Spin-inertia and spin-gravity interactions are the subject of numerous theoretical
\cite{hehlni,hehl,audretsh,huang,mashhoon,mashhoon1,singh,obukhov} and experimental
efforts \cite{ni,silverman,versuve,wineland,ritter}. At the same time precise Earth-bound
and near space experimental tests of fundamental theories require that inertial effects
be identified with great accuracy. It is shown below that spin-rotation coupling is
particularly important in precise tests of fundamental theories and in certain types of
neutrino oscillations. Surprisingly, particle accelerators may be also called to play a
role in these investigations \cite{caill}.

\section{Wave equations}

The quantum phases induced by inertia and gravity are derived  in this section for
Schroedinger, Klein-Gordon, Maxwell and Dirac equations. Some applications are given in
Section 3.

\subsection{The Schroedinger equation}

Starting from the action principle ($\hbar = c = 1$)
\begin{equation}\label{2.1}
 S = - m \int{ds} = - m
\int{\sqrt{g_{\mu\nu}\dot{x}^{\mu}\dot{x}^{\nu}} dx^{0}}{,}
\end{equation}
where $\dot{x}^{\mu }=dx^{\mu}/dx^{0}$, one arrives at the Lagrangian
\begin{equation}\label{2.2}
L = -m(g_{ij}\dot{x}^{i}\dot{x}^{j} + 2 g_{i0}\dot{x}^{i} + g_{00})^{1/2}{,}
\end{equation}
where $i,j=1,2,3$. From $L$ one obtains
\begin{equation}\label{2.3}
p_{i}=\partial{L}/\partial{\dot{x}^{i}}=-m(g_{ij}\dot{x}^{j}+g_{i0})(g_{lk}\dot{x}^{l}\dot{x}^{k}+
2g_{k0}\dot{x}^{k}+g_{00})^{-1/2} {.}
\end{equation}
Substituting (\ref{2.3}) into $H=p_{i}\dot{x}^{i}-L$ one finds
\begin{equation}\label{2.4}
H=m(g_{i0}\dot{x}^{i}+g_{i0})(g_{lk}\dot{x}^{l}\dot{x}^{k}+
2g_{k0}\dot{x}^{k}+g_{00})^{-1/2}{.}
\end{equation}
In the WFA $g_{\mu\nu}\simeq\eta_{\mu\nu}+\gamma_{\mu\nu},
 g^{\mu\nu}\simeq\eta^{\mu\nu}-\gamma^{\mu\nu},  g^{ij}g_{jk}\simeq\delta^{i}_{k}$.
 From (\ref{2.3}) one then gets
\begin{equation}\label{2.5}
g^{ij}p_{j}\simeq-m(\dot{x}^{i}+g^{ij}g_{j0})(g_{lk}\dot{x}^{l}\dot{x}^{k}+
2g_{k0}\dot{x}^{k}+g_{00})^{-1/2}{.}
\end{equation}
Eq.(\ref{2.5}) can be solved for $-m\dot{x}^{i}/(g_{lk}\dot{x}^{l}\dot{x}^{k}+
2g_{k0}\dot{x}^{k}+g_{00})^{-1/2}$ and gives
\begin{equation}\label{2.6}
\dot{x^{j}}= -\frac{1}{m} (g_{lk}\dot{x}^{l}\dot{x}^{k}+
2g_{k0}\dot{x}^{k}+g_{00})^{1/2}g^{jk}p_{k}-g^{jk}g_{k0} {.}
\end{equation}
On using Eq.(\ref{2.6}), one finds
\begin{equation}\label{2.7}
g_{lk}\dot{x}^{l}\dot{x}^{k}+
2g_{k0}\dot{x}^{k}+g_{00}=(g_{00}-g^{il}g_{i0}g_{l0})/(1-1/m^{2} g^{lk}p_{l}p_{k}){.}
\end{equation}
By substituting (\ref{2.6}) and (\ref{2.7}) into (\ref{2.4}), one obtains
\begin{equation}\label{2.8}
H\simeq\sqrt{p^{2}+m^{2}}(1+1/2 \gamma_{00})+1/2
\gamma^{ij}p_{i}p_{j}/\sqrt{p^{2}+m^{2}}-p^{l}\gamma_{l0} {.}
\end{equation}
In the presence of electromagnetic fields and in the low velocity limit, the Hamiltonian
(\ref{2.8}) leads to the Schroedinger equation \cite{pap1}
\begin{equation}\label{2.9}
i\partial{\psi{(x)}}/\partial{t}=[1/2m (p_{i}-e A_{i}+m \gamma_{0i})^{2}-e A_{0}+1/2m
\gamma_{00}]\psi{(x)} {.}
\end{equation}
The WFA does not fix the reference frame entirely. The transformations
$x_{\mu}\rightarrow x_{\mu}+\xi_{\mu}$ are still allowed and lead to the "gauge"
transformations $\gamma_{\mu\nu}\rightarrow\gamma_{\mu\nu}-\xi_{\mu,\nu}-\xi_{\nu,\mu}$.
In the stationary case the transformations $\gamma_{00}\rightarrow\gamma_{00},
  \gamma_{0i}\rightarrow\gamma_{0i}-\xi_{0i}$ leave Eq.(\ref{2.9}) invariant. Returning to
  normal units,
  the solution of
the Schroedinger equation is in this case
\begin{equation}\label{2.10}
\psi{(x)}=exp\left\{imc/\hbar \int^{x}{\gamma_{0i} dx^{i}} -
ie/c\hbar\int^{x}{A_{i}dx^{i}}\right\}\psi_{0}{(x)} {,}
\end{equation}
where $\psi_{0}$ is the solution of the field-free Schroedinger equation. If the
electron-lattice interaction is added to Eq.(\ref{2.9}), then the resulting equation can
be applied to the study of BCS superconductors in weak stationary gravitational fields
\cite{dewitt,pap2}. This is desirable because BCS superconductors behave in many respects
as non viscous fluids. They also exhibit quantization on a macroscopic scale and appear
ideally suited to magnify small physical effects.

\subsection{The Klein-Gordon Equation}

A well-known form of the the fully covariant Klein-Gordon equation is
\begin{equation}\label{2.11}
(g^{\mu\nu}\nabla_{\mu}\nabla_{\nu}-m^2)\Phi{(x)}=0 {,}
\end{equation}
where $\nabla_{\mu}$ represents covariant differentiation. To first order in the WFA,
Eq.(\ref{2.11}) becomes
\begin{equation}\label{2.12}
[(\eta^{\mu\nu}-\gamma^{\mu\nu})\partial_{\mu}\partial_{\nu}-(\gamma^{\alpha\mu}-1/2
\gamma_{\sigma}^{\sigma}\eta^{\alpha\mu}),_{\mu}\partial_{\alpha}]\phi{(x)}=0 {.}
\end{equation}
Eq.(\ref{2.12}) has the exact solution \cite{caipap1,caipap2}
\begin{equation}\label{2.13}
\Phi(x)=exp\left\{-i \Phi_{g}\right\} \phi_{0}(x)=(1-i \Phi_{g}) \phi_{0}(x) {,}
\end{equation}
where $\phi_{0}(x)$ is the solution of the field-free equation in Minkowski space, and
\begin{eqnarray}\label{2.14}
i \Phi_{g} \phi_{0}&=&[\frac{1}{4}\int_{P}^{x}dz^{\lambda}
(\gamma_{\alpha\lambda,\beta}(z)-\gamma_{\beta\lambda,\alpha}(z))[(x^{\alpha}-z^{\alpha})
\partial^{\beta}-(x^{\beta}-z^{\beta})\partial^{\alpha}]-\nonumber \\
& &
\frac{1}{2}\int_{P}^{x}dz^{\lambda}\gamma_{\alpha\lambda}(z)\partial^{\alpha}]\phi_{0}{.}
\end{eqnarray}
Eq.(\ref{2.14}) is related to Berry's phase \cite{caipap3}. It is easy to prove by direct
substitution that (\ref{2.13}) is a solution of (\ref{2.12}). In fact
\begin{eqnarray}\label{2.15}
i
\partial_{\mu}(\Phi_{g}\phi_{0})&=&\frac{1}{4}\int^{x}_{P}dz^{\lambda}
(\gamma_{\alpha\lambda,\beta}(z)-
\gamma_{\beta\lambda,\alpha}(z))[\delta_{\mu}^{\lambda}\partial^{\beta}-\delta_{\mu}^{\beta}
\partial^{\alpha}]\phi_{0}(x)+\nonumber\\
& & \frac{1}{4}\int_{P}^{x}dz^{\lambda}(\gamma_{\alpha\lambda,\beta}(z)-
\gamma_{\beta\lambda,\alpha}(z)) [(x^{\alpha}-z^{\alpha})\partial^{\beta}-\nonumber\\
& &
(x^{\beta}-z^{\beta})\partial^{\alpha}]
\partial_{\mu}\phi_{0}(x)-
\frac{1}{2}\int_{P}^{x}dz^{\lambda}
\gamma_{\alpha\lambda}(z)\partial^{\alpha}\partial_{\mu}\phi_{0}(x)-\nonumber\\
& &
\frac{1}{2} \gamma_{\alpha\mu}(x)
\partial^{\alpha}\phi_{0}(x) {,}
\end{eqnarray}
from which one gets
\begin{equation}\label{2.16}
i\partial_{\mu}\partial^{\mu}(\Phi_{g}\phi_{0})=im^2
\Phi_{g}\phi_{0}-\gamma_{\mu\alpha}\partial^{\mu}\partial^{\alpha}\phi_{0}-
(\gamma^{\beta\mu}-
\frac{1}{2}\gamma_{\sigma}^{\sigma}\eta^{\beta\mu}),_{\mu}\partial_{\beta}\phi_{0}{.}
\end{equation}
The result is proven by substituting (\ref{2.16}) into (\ref{2.12}). For a closed path in
space-time one finds
\begin{equation}\label{2.17}
i\Delta\Phi_{g}\phi_{0}=\frac{1}{4}\oint{R_{\mu\nu\alpha\beta}L^{\alpha\beta}
d\tau^{\mu\nu}}\phi_{0}{,}
\end{equation}
where $L^{\alpha\beta}$is the angular momentum of the particle of mass $m$ and
$R_{\mu\nu\alpha\beta}$ is the linearized Riemann tensor. The result found is therefore
manifestly gauge invariant.

Unlike the case of the Schroedinger equation discussed above, the gravitational fields
considered in this section need not be stationary.

Since Eq.(\ref{2.13}) is also a solution of the Landau-Ginzburg equation \cite{caipap1},
the present results may be applied to the description of charged and neutral superfluids
and Bose-Einstein condensates.

Applications of (\ref{2.13}) to the detection of gravitational waves can be found in the
literature \cite{caipap1,thompson}.

\subsection{Maxwell equations}

Consider now Maxwell equations
\begin{equation}\label{2.18}
\nabla_{\nu}\nabla^{\nu}A_{\mu}-R_{\mu\sigma}A^{\sigma}=0 {,}
\end{equation}
where the electromagnetic field $A_{\mu}$ satisfies the condition$\nabla_{\mu}A^{\mu}=0$.
If the second term in Eq.(\ref{2.18}) is negligible, then Maxwell equations in the WFA
are
\begin{equation}\label{2.19}
\nabla_{\nu}\nabla^{\nu}A_{\mu}\simeq(\eta^{\sigma\alpha}-\gamma^{\sigma\alpha})
A_{\mu,\alpha\sigma}+ R_{\mu\sigma}A^{\sigma}-(\gamma_{\sigma\mu,\nu}+
\gamma_{\sigma\nu,\mu}-\gamma_{\mu\nu,\sigma})A^{\sigma,\nu}=0 {,}
\end{equation}
where use has been made of the Lanczos-DeDonder gauge condition
\begin{equation}\label{2.20}
\gamma_{\alpha\nu},^{\nu}-\frac{1}{2}\gamma_{\sigma}^{\sigma},_{\alpha}=0 {.}
\end{equation}
Eq.(\ref{2.19}) has the solution \cite{caipap2}
\begin{eqnarray}\label{2.21}
A_{\mu}(x)&=&a_{\mu}(x)-\frac{1}{4}\int_{P}^{x}dz^{\lambda}(\gamma_{\alpha\lambda,\beta}(z)-
\gamma_{\beta\lambda,\alpha}(z))[(x^{\alpha}-z^{\alpha})\partial^{\beta}a_{\mu}(x)
-\nonumber \\
& &
(x^{\beta}-z^{\beta})\partial^{\alpha}a_{\mu}(x)]+\frac{1}{2}\int_{P}^{x}dz^{\lambda}
\gamma_{\alpha\lambda}(z)
\partial^{\alpha}a_{\mu}(x)+\nonumber \\
& &
\frac{1}{2}\int_{P}^{x}dz^{\lambda}(\gamma_{\beta\mu,\lambda}(z)+\gamma_{\beta\lambda,\mu}(z)-
\gamma_{\mu\lambda,\beta}(z))a^{\beta}(x) {,}
\end{eqnarray}
where $\partial_{\nu}\partial^{\nu}a_{\mu}=0$ and $\partial^{\nu}a_{\nu}=0$.
Eq.(\ref{2.21}) can also be written in the form $A_{\mu}=exp(-i\xi)a_{\mu}$, where
\begin{eqnarray}\label{2.22}
 \xi&=&-\frac{1}{4}\int_{P}^{x}dz^{\lambda}(\gamma_{\alpha\lambda,\beta}(z)-
\gamma_{\beta\lambda,\alpha}(z))J^{\alpha\beta}+\nonumber\ \\ & &
\frac{1}{2}\int_{P}^{x}dz^{\lambda}\gamma_{\alpha\lambda}(z)k^{\alpha}-
\frac{1}{2}\int_{P}^{x}dz^{\lambda}\gamma_{\alpha\beta,\lambda}(z)T^{\alpha\beta}{,}
\end{eqnarray}
 $J^{\alpha\beta}=L^{\alpha\beta}+S^{\alpha\beta}$ is the total angular momentum,
$(S^{\alpha\beta)\mu\nu}=-i(g^{\mu\alpha}g^{\nu\beta}\\-g^{\mu\beta}g^{\nu\alpha})$ is
the spin-1 operator, $(T^{\alpha\beta})^{\mu\nu}\equiv
-i\frac{1}{2}(g^{\mu\alpha}g^{\nu\beta}+g^{\mu\beta}g^{\nu\alpha})$ and $k^{\alpha}$ is
the momentum of the free photon. All spin effects are therefore contained in the
$S^{\alpha\beta}$ and $T^{\alpha\beta}$ terms. For a closed path one can again find
Eq.(\ref{2.17}).

\subsection{The generally covariant Dirac equation}

Some of the most precise experiments in physics involve spin-1/2 particles. They are very
versatile tools that can be used in a variety of experimental situations and energy
ranges while still retaining essentially a non-classical behaviour. Within the context of
general relativity, De Oliveira and Tiomno \cite{de} and Peres \cite{peres} conducted
comprehensive studies of the fully covariant Dirac equation. The latter takes the form
\begin{equation}\label{2.23}
[i\gamma^{\mu}(x)D_{\mu}-m]\Psi(x)=0 {,}
\end{equation}
where $D_{\mu}=\nabla_{\mu}+i\Gamma_{\mu}$, $\{\gamma^{\mu}(x),\gamma^{\nu}(x)\}=2
g^{\mu\nu}(x)$, and the spin connection $\Gamma^{\mu}$ is defined by
$D_{\mu}\gamma_{\nu}(x)=
\nabla_{\mu}\gamma_{\nu}(x)+i[\Gamma_{\mu}(x),\gamma_{\nu}(x)]=0$. By using the
definitions $ \Psi(x)=S \tilde{\Psi}(x) $, $
S=exp(-i\int_{P}^{x}dz^{\lambda}\Gamma_{\lambda}(z)) $ and $
\tilde{\gamma}^{\mu}(x)=S^{-1}\gamma^{\mu}(x)S$, in (\ref{2.23}) one finds
\begin{equation}\label{2.24}
 [i\tilde{\gamma}^{\mu}(x)\nabla_{\mu}-m]\tilde{\Psi}=0 {.}
 \end{equation}
By substituting $ \tilde{\Psi}=[-i\tilde{\gamma}^{\alpha}(x)\nabla_{\alpha}-m]\psi'$ into
(\ref{2.24}) , one obtains
\begin{equation}\label{2.25}
(g^{\mu\nu}\nabla_{\mu}\nabla_{\nu}+m^2)\psi'=0
 \end{equation}
 which, as shown above, has
the WFA solution $\psi'=exp(-i\Phi_{g})\psi_{0}$, where $\psi_{0}$ is a solution of the
Dirac equation in Minkowski space. It is again possible to show that for a closed path
the total phase difference suffered by the Dirac wave function is gauge invariant and is
given by $ -\frac{1}{4}\oint R_{\mu\nu\alpha\beta}J^{\alpha\beta}d\tau^{\mu\nu} $, where
the total angular momentum is now $ J^{\alpha\beta}=L^{\alpha\beta}+ \sigma^{\alpha\beta}
$, $ \sigma^{\alpha\beta}=-\frac{1}{2}[\gamma^{\alpha},\gamma^{\beta}] $ and
$\gamma^{\beta}$ represents a usual, constant Dirac matrix \cite{caipap2}.

\section{Applications}

Several applications of solutions (\ref{2.10}) and (\ref{2.13})to superconductors,
gyroscopy and interferometry can now be discussed.

\subsection{Superconductors}

By comparing the Schroedinger equation for superconductors in electromagnetic fields with
(\ref{2.9}) one can immediately draw the following conclusions \cite{dewitt,pap2}.

i)$\overrightarrow{\nabla}(A_{0}-\frac{1}{2}\frac{m c^2}{e}\gamma_{00})=0$. This means
that the gravitational field generates an electric field inside the superconductor,
contrary to the gravity-free case ($\gamma_{00}=0$) that yields $\vec{E}=0$. In principle
one could therefore detect a gravitational field by means of the electric field it
produces inside the superconductor. If the field is Newtonian, then $ \vec{E}=
\frac{mg}{e}$ which is the field Schiff and Barnhill \cite{schiff} predicted gravity
would produce inside normal conductors.

ii)$B_{i}+ \frac{mc^2}{e}\varepsilon_{ijk}\partial^{j}\gamma^{0i}=0$ well inside the
superdonductor where $B_{i}$ is known to vanish in the absence of gravitational fields.

iii) The total flux $ \oint (A_{i}-\frac{mc^2}{e}\gamma_{0i})dx^{i}=n \frac{\hbar c}{2e}$
is quantized, rather than just the flux of $B_{i}$. This again signifies that
$\gamma_{0i}$ could be measured if the magnetic field it generates were sufficiently
large. When the superconductor rotates $\gamma_{0i}=(\frac{\overrightarrow{\omega}\times
\overrightarrow{r}}{c})_{i}$, one finds $
\overrightarrow{B}=\frac{2mc}{e}\overrightarrow{\omega}$ which is the London moment of
rotating superconductors. This result offers tangible evidence that inertia interacts
with a quantum system in ways that are compatible with Einstein's views down to lengths
of the order of $10^{-4}cm$.

These conclusions only apply to $stationary$ gravitational fields. Other examples of
gravity-induced electric and magnetic fields are discussed in the literature \cite{pap2}.

\subsection{Rotation}

Consider for simplicity a square interferometer ABCD of side $l$ in the (xy)-plane,
rotating with angular velocity $\omega$ about the z-axis. The emission and interference
of spinless particles of mass $m$ take place at A and C respectively. Using the metric
\begin{equation}\label{2.26}
ds^2=(1-\omega^2 \frac{x^2+y^2}{c^2})(dx^{0})^2+\frac{2\omega}{c}(y dx-x
dy)dx^{0}-dx^2-dy^2-dz^2 {,}
\end{equation}
and indicating by $\wp_{1}$ the path ABC and by $\wp_{2}$ the path ADC, the non-vanishing
contributions to $i\Phi_{g}\phi_{0}$ given by Eq.(\ref{2.14}) are
\begin{eqnarray}\label{2.27}
\Delta\chi&=&-\frac{1}{2}\int_{A,\wp_{1}}^{C}dz^{\lambda}\gamma_{\alpha\lambda}(z)k^{\alpha}+
\frac{1}{2}\int_{A,\wp_{2}}^{C}dz^{\lambda}\gamma_{\alpha\lambda}(z)
k^{\alpha}=\nonumber\\ & &
-\frac{1}{2}\int(dz^{0}\gamma_{10}k^{1}+dz^{0}\gamma_{20}k^{2}+dz^{1}\gamma_{01}k^{0}+
dz^{2}\gamma_{02}k^{0})=\nonumber\\ & & \frac{\omega}{2c}
[-\int_{0,\wp_{1}}^{\frac{lc}{v}}dz^{0}y k^{1}+
\int_{\frac{lc}{v}}^{\frac{2lc}{v}}dz^{0}x k^{^2}-\int_{0,\wp_{2}}^{\frac{lc}{v}}dz^{0}x
k^{2}+\nonumber\\ & &\int_{\frac{lc}{v}\wp_{2}}^{\frac{2lc}{v}}dz^{0} y k^{1}]- k^{0}
[\int_{0,\wp_{1}}^{l}y dx-\int_{0,\wp_{1}}^{l}x dy+\int_{0,\wp_{2}}^{l}x dy-
\int_{0,\wp_{2}}^{l}y dx]=\nonumber\\ & & \frac{\omega l^2}{c}(k \frac{c}{v}+k^{0}) {.}
\end{eqnarray}
For non-relativistic particles $ k^{0}\sim\frac{mc}{\hbar}(1+\frac{v^2}{2c^2})$, $
k\sim\frac{mv}{\hbar}$ and the result is
$\Delta\chi\sim\frac{2ml\omega^2}{\hbar}(1+\frac{v^2}{8c^2})$. The first term agrees with
the results of several relativistic and non-relativistic approximations. In general one
obtains from Eq.(\ref{2.14})
\begin{equation}\label{2.28}
 \Delta\chi=(\frac{2m}{\hbar}+\frac{\hbar
k^2}{2mc^2})\vec{\omega}\cdot\vec{a} {,}
\end{equation}
where $ \vec{a}$ represents the area of the interferometer oriented along its normal
\cite{caipap1}. It therefore appears that gyroscopy is completely controlled by the
quantum phase (\ref{2.28}). One also finds that $
\frac{(\Delta\chi)_{ph}}{(\Delta\chi)_{part}}=\frac{{\lambda}_{c}}{(\lambda)_{ph}} $,
where $ \lambda_{c} $ is the Compton wavelength of the particle circulating in the
interferometer. This ratio indicates that particle interferometers are more sensitive
than photon interferometers for particle masses $m>\frac{h\nu_{ph}}{c^{2}}$.

On applying (\ref{2.22}) to photons, one finds that the time integral part of $ \xi$
yields
\begin{eqnarray}\label{2.29}
-\frac{1}{4}\int_{P}^{x}dz^{0}(\gamma_{\alpha0,\beta}-\gamma_{\beta0,\alpha})S^{\alpha\beta}-
\frac{1}{2}\int_{P}^{x}dz^{0}\gamma_{\alpha\beta,0}T^{\alpha\beta}&=&\nonumber\ \\
-\frac{1}{2}\int_{P}^{x}dz^{0}\gamma_{i0,j} S^{ij}= \int dt \omega S_{z}
\end{eqnarray}
which represents the spin-rotation coupling , or Mashhoon effect, for photons
\cite{mashhoon,mashhoon1,caipap2}.

\subsection{Gravitational red-shift}

Two light sources of the same frequency are at distances $ r_{A} $ and $ r_{B} $ from the
origin at the initial time $ x_{1}^{0}$. They are compared at $ r_{A}$ at the later time
$x_{2}^{0}$. Neglecting spin effects, the phase difference can be simply obtained from
 (\ref{2.14}) using the closed space-time path in the $ (r,x^{0})$-plane
with vertices at $ (r_{A}, x_{1}^{0}), (r_{B}, x_{1}^{0}), (r_{B}, x_{2}^{0}), (r_{A},
x_{2}^{0}) $. The gravitational field is represented by $ \gamma_{00}(r)=2 \varphi(r) $,
where $ \varphi(r) $ is the Newtonian potential. One finds
\begin{eqnarray}\label{2.30}
\Delta\chi &=&
\frac{1}{2}\int_{x_{1}^{0}}^{x_{2}^{0}}dz^{0}[\gamma_{\alpha0,\beta}(r_{B})-
\gamma_{\beta0,\alpha}(r_{B})] (x^{\alpha}-z^{\alpha})k^{\beta}+\nonumber\\ & &
\frac{1}{2}\int_{x_{2}^{0}}^{x_{1}^{0}}dz^{0} \gamma_{\alpha
0,\beta}(r_{A})(x^{\alpha}-z^{\alpha})k^{\beta}-\frac{1}{2}\int_{x_{1}^{0}}^{x_{2}^{0}}dz^{0}
\gamma_{\alpha0}(r_{B})k^{\alpha}- \nonumber\\ & &
\frac{1}{2}\int_{x_{2}^{0}}^{x_{1}^{0}}dz^{0}\gamma_{\alpha0}(r_{A})k^{\alpha}=\nonumber\\
& &
-\frac{k^{0}}{2}(x_{2}^{0}-x_{1}^{0})[\gamma_{00}(r_{B})-\gamma_{00}(r_{A})]-\nonumber\\
& & \frac{k^{0}}{4}(x_{2}^{0}-x_{1}^{0})^2[\gamma_{00,1}(r_{B})+\gamma_{00,1}(r_{A})] {.}
\end{eqnarray}
The first term gives the usual red-shift formula $
(\frac{\Delta\nu}{\nu})_{1}=-\frac{1}{c^2}[\varphi(r_{B})-\varphi(r_{A})] $. The second
term yields the additional correction $ (
\frac{\Delta\nu}{\nu})_{2}=-\frac{x_{2}^{0}-x_{1}^{0}}{4}[\gamma_{00,1}(r_{B})+
\gamma_{00,1}(r_{A})]$. In an experiment of the type carried out by Pound and Rebka the
ratio of the two terms is $(\frac{\Delta\nu}{\nu})_{2}/(\frac{\Delta\nu}{\nu})_{1}\simeq
\frac{2l}{R_{\oplus}}$, where $l=r_{B}-r_{A}$. The second term
$(\frac{\Delta\nu}{\nu})_{2}$ should therefore be measurable for sufficiently high values
of $l$.

\subsection{Schwarzschild metric}

If Earth is assumed perfectly spherical and homogeneous and rotation is neglected, then
its gravitational field can be described by the Schwarzschild metric\cite{fang}
\begin{equation}\label{2.31}
ds^2=(1-\frac{2M_{\oplus}}{r})(dx^{0})^2-(1-\frac{2M_{\oplus}}{r})^{-1}dr^2-r^2d\theta^2-
r^2\sin\theta^2d\varphi^2 {.}
\end{equation}
Assuming, for simplicity that a square interferometer of side $ l $ is placed in a
vertical plane at one of the poles and that particle emission and interference occur at
opposite corners, one finds
\begin{equation}\label{2.32}
\Delta\chi=\frac{GM_{\oplus}l^2m}{R_{\oplus}^2\hbar v}\left(1+\frac{3 v^2}{2 c^2}-\frac{3
l}{2 R_{\oplus}}-\frac{3v^2l}{4c^2R_{\oplus}}\right) {.}
\end{equation}
The first term in Eq.(\ref{2.32}) is the term observed with a neutron interferometer in
the well known COW experiment \cite{colella}. If $v\sim 10^{-5} c$, then the De Broglie's
wavelength for neutrons is $\sim 10^{-8}cm$. General relativity appears therefore to be
valid down to lengths of this order of magnitude. The ratio of the terms on the r.h.s. of
(\ref{2.32}) is $ 1:10^{-10}:10^{-7}:10^{-17}$. The last term is extremely small and may
be neglected. Of the remaining terms, the second represents a special relativistic
correction, which is smaller than the general relativity effect represented
 by the third term. The ratio of the latter to the magnitude of the Sagnac effect also is
 $\sim 10^{-7}$ and appears difficult to observe at present.

\subsection{Lense-Thirring field of Earth}

The non-vanishing components of $\gamma_{\mu\nu}$ are in this instance \cite{lense}
\begin{eqnarray}\label{2.33}
\gamma_{00}&=&\gamma_{11}=\gamma_{22}=\gamma_{33}=\frac{2GM_{\oplus}}{c^2r}\nonumber \\
\gamma_{01}&=& \frac{4GM_{\oplus}\omega a^2(y+y')}{5c^3r^3},
\gamma_{02}=\frac{4GM_{\oplus}\omega a^2(x+x')}{5c^3r^3} {,}
\end{eqnarray}
where $r^2=(x+x')^2+(y+y')^2+(z+z')^2$, Earth is again assumed spherical and homogeneous,
$\omega$ its angular velocity about the $z'$-axis, and $(x',y',z')$ are the coordinates
at the point $A$ at which the interferometer beam is split in the coordinate system with
origin at the centre of the Earth. Interference occurs at the opposite vertex $C$. The
frame $z^{\mu}$ has origin at $A$, is at rest relative to $z'^{\mu}$ and the plane of the
interferometer is chosen for simplicity to coincide with the $(x,y)$-plane and parallel
to the $(x',y')$-plane. The time at which the particle beam is split at $A$ is
$z^{0}=z^{'0}=0 $ \cite{caipap4}. If, in particular, $A$ coincides with a pole, then
$\Delta\chi=\frac{2G}{c^2R^3}J_\oplus\frac{ml^2}{\hbar}$, where
$J_{\oplus}=\frac{2M_{\oplus}R_{\oplus}^2\omega}{5}$ is the angular momentum of Earth.
Taking into account that the precession frequency of a gyroscope in orbit is $
\Omega=\frac{GJ_{\oplus}}{2c^2R_{\oplus}^3}$, one can also write $ \Delta\chi=\Omega
\Pi$, where $ \Pi=\frac{4ml^2}{\hbar}$ replaces the period of a satellite in the
classical calculation. Its value, $\Pi\sim 1.4\times10^{8}s $ for neutron interferometers
with $ l\sim10^2cm$, is rather high and yields $\Delta\chi\sim10^{-7}rad$. This suggests
that the development and use of heavy particle interferometers would be particularly
advantageous in attempts to measure the Lense-Thirring effect.

\section{Helicity precession of fermions}

Consider the line element $ ds^{2}=g_{\mu\nu}(x)dx^{\mu}dx^{\nu} $ and the set of tangent
vectors $ {\vec{e}_{\mu}}={\partial_{\mu}\vec{P}} $ that forms the coordinate basis that
spans the manifold $
g_{\mu\nu}(x)=\partial_{\mu}\vec{P}\cdot\partial_{\nu}\vec{P}\equiv\vec{e}_{\mu}\cdot\vec{e}_{\nu}$
. The principle of equivalence ensures the existence of an orthonormal tetrad frame $
{{\vec{e}_{\hat{\mu}}}}={\partial_{\hat{\mu}}\vec{P}}$ such that for a local tangent
space defined at any given point of space-time $
\eta_{\hat{\mu\nu}}=\vec{e}_{\hat{\mu}}\cdot\vec{e}_{\hat{\nu}}$. The principle
underlying the tetrad formalism therefore requires that for a sufficiently small region
of space-time $ \vec{e}_{\mu}$ be mapped onto $ \vec{e}_{\hat{\mu}} $ using a set of
projection functions $ e_{\hat{\mu}}^{\nu}$ and their inverses $ e_{\hat{\mu}}^{\nu}$
such that
\begin{eqnarray}\label{2.34}
\vec{e}_{\hat{\mu}}&=& e_{\hat{\mu}}^{\nu}\vec{e}_{\nu},
 \vec{e_{\mu}}=e_{\mu}^{\hat{\nu}},
 \eta_{\hat{\mu\nu}}=e_{\hat{\mu}}^{\mu}e_{\hat{\nu}}^{\nu}g_{\mu\nu}(x),\nonumber\\
 e_{\hat{\mu}}^{\nu}e_{\nu}^{\hat{\alpha}}&=&\delta_{\hat{\mu}}^{\hat{\alpha}},
 e_{\mu}^{\hat{\nu}}e_{\hat{\nu}}^{\alpha}=\delta_{\mu}^{\alpha} {.}
 \end{eqnarray}
  When $\vec{e}_{\mu}$
refers to an observer with acceleration $ \vec{a}$ rotating with angular velocity
$\vec{\omega}$ , one finds \cite{hehlni}
\begin{equation}\label{2.35}
ds^{2}=[(1+\vec{a}\cdot\vec{x})^{2}+( \vec{\omega}\cdot\vec{x}) ^{2}-\omega^{2}x^{2}]
dx_{0}^{2}-2dx_{0}d\vec{x}\cdot( \vec{\omega}\times\vec{x}) -d\vec{x}\cdot d\vec{x}{,}
\end{equation}
while
\begin{eqnarray}\label{2.36}
\vec{e}_{\hat{0}}&=&(1+\vec{a}\cdot\vec{x})^{-1}[\vec{e}_{0}-(
\vec{\omega}\times\vec{x})^{k}\vec{e}_{k}], \vec{e}_{\hat{i}}=\vec{e}_{i},\nonumber \\
e_{\hat{0}}^{0}&=&(1+\vec{a}\cdot\vec{x})^{-1},
e_{\hat{0}}^{k}=-(1+\vec{a}\cdot\vec{x})^{-1}(\vec{\omega}\times\vec{x}^{k}),\nonumber \\
e_{0}^{\hat{0}}&=&1+\vec{a}\cdot\vec{x}, e_{0}^{\hat{k}}=( \vec{\omega}\times\vec{x})
^{k}, e_{i}^{\hat{0}}=0, e_{i}^{\hat{k}}=\delta_{i}^{k} {.}
\end{eqnarray}
Also, from $ D_{\mu}\gamma_{\nu}(x)=0 $ and $ \gamma_{\mu}(x) =
e_{\mu}^{\hat{\mu}}\gamma_{\hat{\mu}}$, where $ \gamma_{\hat{\mu}}$ represents the usual
Dirac matrices, one finds $
\Gamma_{\mu}(x)=\frac{1}{4}\sigma^{\hat{\alpha\beta}}\Gamma_{\hat{\alpha\beta\mu}}
e_{\mu}^{\hat{\mu}}$. The Ricci coefficients are $
\Gamma_{\hat{\nu\alpha\beta}}=\frac{1}{2}(
 C_{\hat{\nu\alpha\beta}}+C_{\hat{\alpha\beta\nu}}-C_{\hat{\beta\nu\alpha}}) $ and $
 C_{\hat{\nu\alpha\beta}}=\eta_{\mu\nu}e_{\hat{\alpha}}^{\alpha}e_{\hat{\beta}}^{\beta}(
 \partial_{\alpha}e_{\beta}^{\hat{\mu}}-\partial_{\beta}^{\hat{\mu}}) $. It also follows that $
 \Gamma_{0}=-\frac{1}{2}a_{i}\sigma^{0i}-\frac{1}{2}\vec{\omega}\cdot\vec{\sigma},
 \Gamma_{i}=0 $, with $ \sigma^{0i}=\frac{1}{2}[ \gamma^{0},\gamma^{i}] $. The Hamiltonian
 is obtained by isolating the time derivative in the Dirac equation. The risult is
 \begin{eqnarray}\label{2.37}
 H&=&\vec{\alpha}\cdot\vec{p}+m \beta+V(x)\nonumber\\
 V(x)&=&\frac{1}{2}[(\vec{a}\cdot\vec{x})( \vec{p}\cdot\vec{\alpha})+(
 \vec{p}\cdot\vec{\alpha})( \vec{a}\cdot\vec{x})]+m\beta(
 \vec{a}\cdot\vec{x})-\vec{\omega\cdot}( \vec{L}+\frac{\vec{\sigma}}{2}) {,}
 \end{eqnarray}
 where $ \vec{L}$ is the orbital angular momentum and $ \vec{\sigma}$ are the usual Pauli
 matrices. The first three terms in $ V(x) $ represent relativistic energy-momentum
 effects. The term $-\vec{\omega}\cdot\vec{L}$ is a Sagnac-type effect. The last term,
 $-\frac{1}{2}\vec{\omega}\cdot\vec{\sigma}$, is the spin-rotation coupling, or Mashhoon
 effect. The non-relativistic effects can be obtained by applying three successive
 Foldy-Wouthysen transformations to $ H$. One obtains to lowest order
 \begin{equation}\label{2.38}
 H=m\beta+\beta\frac{p^{2}}{2m}+\beta m( \vec{a}\cdot\vec{x})+\frac{\beta}{2m}\vec{p}(
 \vec{a}\cdot\vec{x})\cdot\vec{p}-\vec{\omega}\cdot(
 \vec{L}+\frac{\sigma}{2})+\frac{1}{4m}\vec{\sigma}\cdot( \vec{a}\times\vec{p}){.}
 \end{equation}
 The third term in Eq.(\ref{2.38}) is the energy-momentum effect observed by Bonse and
 Wroblewski \cite{bonse}. The term $ -\vec{\omega}\cdot\vec{L}$ was predicted by Page \cite{page}
 and observed by
 Werner and collaborators \cite{werner}. The term $-\vec{\omega}\cdot\frac{\vec{\sigma}}{2} $ was found by
 Mashhoon. Hehl and Ni \cite{hehlni} re-derived all terms and also predicted the existence  of the fourth
 term (a kinetic energy effect) and of the last term (spin-orbit
 coupling). Equations (\ref{2.37}) and (\ref{2.38}) can also be obtained by isolating the quantum
 phase in the wave function.
 The spin-rotation coupling term deserves a few comments. As discussed by Mashhoon, the
 effect violates the hypothesis of locality, namely that an accelerated observer is
 locally equivalent to an instantaneously comoving observer. This hypothesis is valid for
 classical point-like particles and optical rays and is widely used in relativity. The
 effect also
  violates the equivalence principle because it does not couple universally to matter \cite{mash}.
 No direct experimental verification of the Mashhoon effect has so far been reported,
 though the data given in \cite{versuve} can be re-interpreted as due to the coupling of Earth's
 rotation to the nuclear spins of mercury. The effect is also consistent with a small
 depolarization of electrons in storage rings \cite{bell}. It is shown below that it plays an
 essential role  in measurements of the anomalous magnetic moment,
 or $ g-2$ factor, of the muon.

\subsection{Spin-rotation coupling in muon g-2 experiments}

Precise measurements of the $g-2$ factor involve muons in a storage ring consisting of a
vacuum tube, a few meters in diameter, in a uniform, vertical magnetic field $ \vec{B}$.
Muons on equilibrium orbits within a small fraction of the maximum momentum are almost
completely polarized with spin vectors pointing in the direction of motion. As the muons
decay, the highest energy electrons with spin almost parallel to the momentum, are
projected forward in the muon rest frame and are detected around the ring. Their angular
distribution does therefore reflect the precession of the muon spin along the cyclotron
orbits \cite{bailey,farley}. Let us start from the covariant Dirac equation (\ref{2.23}).
It is convenient to use the chiral representation for the usual Dirac matrices
\begin{eqnarray}\label{2.39}
\gamma^{0}&=&\beta=\left(\matrix{0&-I\cr-I&0\cr}\right){,}
\gamma^{i}=\left(\matrix{0&\sigma^{i}\cr-\sigma^{i}&0\cr}\right) {,}
\alpha^{i}=\left(\matrix{\sigma^{i}&0\cr0&-\sigma^{i}\cr}\right){,}\nonumber\\
\sigma^{0i}&=&i\left(\matrix{\sigma^{i}&0\cr0&-\sigma^{i}\cr}\right)
{,}\sigma^{ij}=\epsilon^{ij}_{k}\left(\matrix{\sigma^{k}&0\cr0&\sigma^{k}\cr}\right)
{,}\gamma^{5}=\left(\matrix{I&o\cr0&-I\cr}\right) {.}
\end{eqnarray}
One must now add to the the Hamiltonian the effect of a magnetic field $ \vec{B}$ on the
total (magnetic plus anomalous) magnetic moment of the particle. Assuming for simplicity
that all quantities in $H$ are time-independent and referring them to a left-handed tern
of axes comoving with the particle in the $x_{3}$-direction and rotating in the
$x_{2}$-direction, one finds
\begin{eqnarray}\label{2.40}
H &=& \alpha^{3}p_{3}+m
\beta+\frac{1}{2}[-a_{1}R(\alpha^{3}p_{3})-(\alpha^{3}p_{3})a_{1}R]+ \beta m
a_{1}R-\vec{\omega}\cdot\vec{L}-\nonumber\\
& & \frac{1}{2}\omega_{2}\sigma^{2}+\mu
B\sigma^{2}\equiv H_{0}+H'  {,}
\end{eqnarray}
where $B_{2}=-B, \mu=(1+\frac{g-2}{2})\mu_{0}$, $\mu_{0}=\frac{e\hbar}{2mc}$ is the Bohr
magneton, $H'=-\frac{1}{2}\omega_{2}\sigma^{2}+\mu\beta\sigma^{2}$ and R is the radius of
the muon's orbit. Electric fields used to stabilize the orbits and stray radial electric
fields can also affect the muon spin. Their effects can be cancelled by choosing an
appropriate muon momentum and will be neglected in what follows.
 Before decay the muon states can be represented as
\begin{equation}\label{2.41}
|\psi(t)>=a(t)|\psi_{+}>+b(t)|\psi_{-}> {,}
\end{equation}
where $|\psi_{+}>$ and $|\psi_{-}>$ are the right and left helicity states of $H_{0}$.
Substituting (\ref{2.41}) into the Schroedinger equation $i\frac{\partial}{\partial
t}|\psi(t)>=H|\psi(t)>$, one obtains
\begin{eqnarray}\label{2.42}
i\frac{\partial}{\partial t}\left( \matrix{a\cr
b\cr}\right)&=&\left(\matrix{<\psi_{+}|(H_{0}+H')|\psi_{+}> &<\psi_{+}|H'|\psi_{-}>\cr
<\psi_{-}|H'|\psi_{+}>&<\psi_{-}|(H_{0}+H')|\psi_{-}>\cr}\right)\left(\matrix{a\cr
b\cr}\right)\nonumber\\ = & &\left(\matrix{E-i\frac{\Gamma}{2}&i(\frac{\omega_{2}}{2}-\mu
B)\cr-i( \frac{\omega_{2}}{2}-\mu B)&E-i\frac{\Gamma}{2}\cr}\right)\equiv
M\left(\matrix{a\cr b\cr}\right) {,}
\end{eqnarray}
where $\Gamma$ represents the width of the muon. Notice that the spin-rotation coupling
is off diagonal in (\ref{2.42}). This is a clear indication that the Mashhoon effect
violates the equivalence principle \cite{mash}. The matrix $ M$ can be diagonalized. Its
eigenvalues are $h_{1}=E-i\frac{\Gamma}{2}+(\frac{\omega_{2}}{2})-\mu B),
h_{2}=E-i\frac{\Gamma}{2}-( \frac{\omega_{2}}{2}-\mu B)$, with the corresponding
eigenvectors
\begin{equation}\label{2.43}
|\psi_{1}>=\frac{1}{\surd2}[ i|\psi_{+}>+|\psi_{-}>];
|\psi_{2}>=\frac{1}{\surd2}[-i|\psi_{+}>+|\psi_{-}>] {.}
\end{equation}
The solution of Eq.(\ref{2.42}) is therefore
\begin{eqnarray}\label{2.44}
|\psi(t)>&=&\frac{1}{\surd2}( e^{-ih_{1}t}|\psi_{1}>+e^{-ih_{2}t}\psi_{2}>)=\nonumber\\&
&\frac{1}{2}[( ie^{-ih_{1}t}-ie^{-ih_{2}t})|\psi_{+}>+(
e^{-ih_{1}t}+e^{-ih_{2}t})|\psi_{-}>]{,}
\end{eqnarray}
where $ |\psi(0)>=|\psi_{-}>$. The spin-flip probability is
\begin{equation}\label{2.45}
P_{\psi_{-}\rightarrow \psi_{+}}=|<\psi_{+}|\psi>|^{2}=\frac{e^{-\Gamma
t}}{2}[1-\cos(2\mu B-\omega_{2})t] {,}
\end{equation}
where the $\Gamma$-term accounts for the observed exponential decrease in electron counts
due to the loss of muons by radioactive decay \cite{farley}. The spin-rotation
contribution to$ P_{\psi_{}\rightarrow \psi_{+}}$ is represented by $\omega_{2}$ which is
the cyclotron angular velocity $ \frac{eB}{m}$. The spin-flip angular frequency is then
\begin{equation}\label{2.46}
\Omega=2\mu
B-\omega_{2}=(1+\frac{g-2}{2})\frac{eB}{m}-\frac{eB}{m}=\frac{g-2}{2}\frac{eB}{m} {,}
\end{equation}
which is precisely the observed modulation frequency of the electron counts
\cite{picasso}(see also Fig. 19 of Ref.\cite{farley}). This result is independent of the
value of the anomalous magnetic moment of the particle. It is therefore the Mashhoon
effect that gives prominence to the $g-2$ term in $\Omega$ by exactly cancelling, in
$2\mu B$, the much larger contribution $\mu_{0}$ that comes from fermions with no
anomalous magnetic moment \cite{paplamb}.

It is perhaps odd that spin-rotation coupling as such has almost gone unnoticed for such
a long time. It is however significant that its effect is observed in an experiment that
has already provided crucial tests of quantum electrodynamics  and a test of Einstein's
time-dilation formula to better than a 0.1 percent accuracy. Recent versions of the
experiment \cite{carey,brown1,brown2} have improved the accuracy of the measurements from
270 ppm to 1.3 ppm. This bodes well for the detection of effects involving spin, inertia
and electromagnetic fields, or inertial fields to higher order.

\section{Neutrino Oscillations}

Neutrino beams produced in weak interactions may be considered as a superposition of
different mass eigenstates. As a beam propagates, different components of the beam evolve
differently so that the probability of finding different eigenstates in the beam varies
with time. The consequences of this can be explored in a number of cases.

\subsection{Neutrino helicity oscillations}

Let us consider a beam of high energy neutrinos. If the neutrino source rotates, the
effective Hamiltonian for the mass eigenstates can be written as $
H_{e}=(p^{2}+m^{2})^{\frac{1}{2}}+\Gamma_{0}\approx p+ \frac{m^{2}}{2E}-
\frac{1}{2}\vec{\omega}\cdot\vec{\sigma}$. For simplicity, consider a one-generation of
neutrino that can now be written as a superposition of $\nu_{L}$ and $\nu_{R}$
 in the form
\begin{equation}\label{2.47}
|\nu(t)>=a_{L}(t)|\nu_{L}>+b_{R}(t)|\nu_{R}> {.}
\end{equation}
It is well known that the standard model contemplates only the existence of $\nu_{L}$,
while $\nu_{R}$ is considered sterile and therefore unobservable. Strictly speaking one
should consider the helicity states $\nu_{\pm}$(that are mass eigenstates) in
(\ref{2.47}), however at high energies $\nu_{L}\simeq \nu_{-}, \nu_{R}\simeq\nu_{+}$.
Assuming that $m_{1}\neq m_{2}$, taking $p_{1}\sim p_{2}$ along the $x_{3}$-axis and
substituting (\ref{2.47}) into the Schroedinger equation that corresponds to $H_{e}$, one
obtains
\begin{equation}\label{2.48}
i\frac{\partial}{\partial t}\left(\matrix{a_{L}\cr
b_{R}\cr}\right)=\left(\matrix{p+\frac{m_{1}^{2}}{2E}&-\frac{\omega_{1}}{2}
-i\frac{\omega_{2}}{2}\cr-\frac{\omega_{1}}{2}+i\frac{\omega_{2}}{2}&p+\frac{m_{2}^{2}}{2}
\cr} \right)\left(\matrix{a_{L}\cr b_{R}\cr} \right) \equiv M_{12}\left(\matrix{a_{L}\cr
b_{R}\cr}\right) {.}
\end{equation}
The eigenvalues of $M_{12}$ are
\begin{equation}\label{2.49}
k_{\mp}=p+\frac{m_{1}^{2}+m_{2}^{2}}{4E}\mp[ ( \frac{\Delta
m^{2}}{2E})^{2}+\omega_{\bot}^{2}] ^{\frac{1}{2}} {,}
\end{equation}
where $\Delta m^{2}\equiv m_{1}^{2}-m_{2}^{2}$, and $\omega_{\bot}^{2}\equiv
\omega_{1}^{2}+\omega_{2}^{2}$. The eigenvectors are
\begin{equation}\label{2.50}
|\nu_{1}>=b_{1}[ \eta_{1}|\nu_{L}>+|\nu_{R}> ], |\nu_{2}>=b_{2}[
\eta_{2}|\nu_{L}>+|\nu_{R}> ] {,}
\end{equation}
where
\begin{equation} \label{2.51}
\eta_{1}=\frac{\omega_{1}+i\omega_{2}}{\Omega+\frac{\Delta m^{2}}{2E}},
 \eta_{2}=\frac{\omega_{1}+i\omega_{2}}{-\Omega+\frac{\Delta m^{2}}{2E}},
\end{equation}
\begin{equation}\label{2.52}
|b_{1}|^{2}= \frac{1}{1+|\eta_{1}|^{2}}, |b_{2}|^{2}=\frac{1}{1+|\eta_{2}|^{2}} {,}
\end{equation}\label{2.53}
and \begin{equation} \Omega=\left[\left(\frac{\Delta
m^{2}}{2E}\right)^{2}+\omega_{\bot}^{2}\right]^{\frac{1}{2}} {.}
\end{equation}
 One therefore finds
\begin{eqnarray}\label{2.54}
|\nu(t)>&=&\frac{b_{1}}{\eta_{1}-\eta_{2}}exp\left[ -i\left(
p+\frac{m_{1}^{2}+m_{2}^{2}}{4E}\right)t\right] \times \nonumber\\& & \left[\left(
e^{i\frac{\Omega}{2}t}\eta_{1}-e^{-i\frac{\Omega}{2}t}\eta_{2}\right)\nu_{L}+2i\sin(
\frac{\Omega t}{2})\nu_{R}\right] {,}
\end{eqnarray}
where the initial condition is $ \nu(0)=\nu_{L}$. One obtains the transition probability
\begin{equation}\label{2.55}
P_{\nu_{L}\rightarrow \nu_{R}}=|<\nu_{R}|\nu(t)>|^{2}=
\frac{\omega_{\bot}^{2}}{2\Omega^{2}}[1-\cos( \Omega t)] {.}
\end{equation}
If the neutrinos have mass, then the magnitude of the transition probability becomes
appreciable if
\begin{equation}\label{2.56}
\omega_{\bot}\geq \frac{\Delta m^{2}}{2E} {.}
\end{equation}
Unlike the flavour oscillations generated by the MSW-mechanism \cite{w,ms} that require
$\Delta m^{2}\neq 0$, helicity oscillations can occur also when $m_{1}=m_{2}$ and
$m_{1}=m_{2}=0$. They are interesting because $\nu_{R}$'s, if they exist, do not interact
with matter and would therefore provide an energy dissipation mechanism with possible
astrophysical implications. The conversion rate that $\nu_{L}\rightarrow \nu_{R}$ is not
large for galaxies and white dwarfs. Assume in fact
 that $\omega_{\bot}\gg \Delta m^{2}/2E$ and that the beam of neutrinos consists of $N_{L}(0)$
 particles at $z=0$. One immediately obtains from (\ref{2.55}) that the relative numbers
 of particles at $z=0$ are \cite{papg}
 \begin{equation}\label{2.57}
 N_{L}(z)=N_{L}(0) \cos^{2}\left(\frac{\omega_{\bot}z}{2c}\right),
 N_{R}(z)=N_{L}(0)\sin^{2}\left(\frac{\omega_{\bot}z}{2c}\right) {.}
 \end{equation}
  One then obtains from
 (\ref{2.57}) $N_{R}\sim 10^{-6}N_{L}(0)$ for galaxies of typical size $L$ such that
 $\omega_{\bot}L\sim 200 km/s$. Similarly, for white dwarfs $ \omega_{\bot}\sim
 1.0 s^{-1} $ and one finds $ N_{R}\sim 10^{-4}N_{L}(0)$. In the case of the Sun $\omega_{\bot}\sim
  7.3\times 10^{-5}-2.4\times 10^{-6}s^{-1}$ and the conversion rate peaks at distances
  $L\sim 10^{15}-4\times 10^{16}cm$, well in excess of the average Sun-Earth distance. Helicity
  oscillations could not therefore explain the solar neutrino puzzle without additional
  assumptions about the Sun's structure \cite{venzo1}.
  For neutron stars, however, the
 dynamics of the star could be affected by this cooling mechanism. In fact neutrinos
 diffuse out of a canonical neutron star in a time $1$ to $10s$, during which they travel
 a distance $3\times 10^{9}cm$ between collisions. At distances $L\sim 5\times 10^{6}cm$
 (the star's
 radius) the conversion rate is $N_{R}\sim 0.5 N_{0}$. Even higher cooling rates may
 occur at higher rotational speeds and prevent the formation of a pulsar.
 These results do not require the existence of a magnetic moment for the neutrino (which
 would also require some mass). Its effect could be taken into account by adding the term
 $ \vec{\mu}\cdot \vec{B}$ to $H_{e}$. In all instances considered, however, magnetic spin-flip
 rates of magnitude comparable to those discussed would require neutrino magnetic moments
 vastly in excess of the value $10^{-19}\mu_{0}\left(\frac{m_{\nu}}{1eV}\right)$
 predicted by the standard $SU(2)\times U(1)$ electroweak theory \cite{venzo2}.

 \subsection{Helicity oscillations in a medium}

 The behaviour of neutrinos in a medium is modified by a potential $V$. When this is taken
 into account, the effective Hamiltonian becomes (after subtracting from
 the diagonal terms a common factor which contributes only to the overall phase)
 \begin{equation}\label{2.58}
 H=p+\frac{m^{2}}{2E}-V-\frac{1}{2}\vec{\omega}\cdot\vec{\sigma} {.}
 \end{equation}
 For simplicity, consider again a one-generation of neutrino and assume that $V$ is constant.
 Applying the digonalization procedure of the previous section to the new Hamilltonian,
 leads to the transition probability
 \begin{equation}\label{2.59}
 P_{\nu_{L}\rightarrow \nu_{R}}=\frac{\omega_{\bot}^{2}}{2\Omega'^{2}}[1-\cos(\Omega't)]
  {,}
\end{equation}
 where $\Omega'=[(V+\frac{\Delta m^{2}}{2E})^{2}+\omega_{\bot}^{2}]
^{\frac{1}{2}}$. One finds from (\ref{2.59}) that spin-flip transitions are strongly
suppressed when $V+\frac{\Delta m^{2}}{2E}>\omega_{\bot}$ and only the $\nu_{L}$
component is present in the beam. If $\omega_{\bot}>V+\frac{\Delta m^{2}}{2E}$, then the
$\nu_{L}$ flux has effective modulation. Resonance occurs at $V=\frac{-\Delta
m^{2}}{2E}$. Consider now the rotating core of a supernova. In this case $V$ can be
relatively large, of the order of several electron volts and corresponds to the
interaction of neutrinos with the particles of the medium. For right-handed neutrinos $V$
vanishes. Assuming that the star does not radiate more energy as $\nu_{R}$'s than as
$\nu_{L}$'s, one finds $L_{\nu_{L}}\sim L_{\nu_{R}}\sim 5\times 10^{53}erg/s$. As the
star collapses, spin-rotation coupling acts on both $\nu_{L}$ and $\nu_{R}$. The
$\nu_{L}$'s become trapped and leak toward the exterior $(l\sim 1.5\times 10^{7}cm)$,
while their interaction with matter is $V\sim 14(\rho/\rho_{c})eV$ and increases
therefore with the medium's density, which at the core is $\rho_{c}\sim 4\times
10^{14}g/cm^{3}$. The $\nu_{R}$'s escape. As $\rho$ increases, the transition
$\nu_{L}\rightarrow \nu_{R}$ is inhibited (off resonance). One also finds $\frac{\Delta
m^{2}}{2E}<\omega_{\bot}$ when $\Delta m^{2}<10^{-5}eV^{2}, E\sim 10
MeV,\omega_{\bot}\sim 6\times 10^{3}s^{-1}$. It then follows from (\ref{2.59}) that
\begin{equation}\label{2.60}
L_{\nu_{L}}\sim L_{\nu_{R}}\sin^{2}(\frac{\omega_{\bot}l}{2c}) {,}
\end{equation}
where $ \frac{\omega_{\bot}l}{2c}\sim \frac{\pi}{2}$. In the time $ \frac{l}{c}\sim
5\times 10^{-4}s$, the energy associated with the $\nu_{R}\rightarrow \nu_{L}$ conversion
is $\sim 2.5\times 10^{50}erg$ which is just the missing energy required to blow  up the
mantle of the collapsing star \cite{papg}.

\subsection{Neutrino flavour oscillations}

Consider a beam of neutrinos of fixed energy $E$ emitted at point $(r_{A},t_{A}) $ of the
$(r,t)$-plane. Assume also that the particles are in a weak flavour eigenstate that is a
linear superposition of mass eigenstates $m_{1}$ and $m_{2}$, with $m_{1}\neq m_{2}$. It
is argued in the literature \cite{ahluw,bhatt} that if interference is observed at the
same space-time point $(r_{B},t_{B})$, then the lighter component must have left the
source at a later time $\Delta t=\frac{r_{B}-r_{A}}{v_{1}}-\frac{r_{B}-r_{A}}{v_{2}}$,
where $v_{1}$ and $v_{2}$ are the velocities of the eigenstates of masses $m_{1}$ and
$m_{2}$ respectively. Because of the difference in travel time $\Delta t$, gravity
induced neutrino flavour oscillations will ensue even though gravity couples universally
to matter. Ignoring spin contributions, the phase difference of the two mass eigenstates
can be calculated from (\ref{2.14}) in a completely gauge invariant way. Assume the
neutrinos propagate in a gravitational field described by the Schwarzschild metric. When
the closed space-time path in (\ref{2.14}) is extended to the triangle $(r_{A},t_{A})$,
$(r_{B},t_{B})$, $(r_{A},t_{A}+\Delta t)$, one obtains
\begin{eqnarray} \label{2.61}
\lefteqn{(i\Delta \Phi_{g}\phi_{0})_{m_{1}}-(i\Delta \Phi_{g}\phi_{0})_{m_{2}}=}\nonumber
\\& & \frac{r_{g}E}{2}[-\frac{v_{1}\Delta t}{2}-\frac{1}{v_{1}}\ln(
\frac{r_{B}}{r_{A}+v_{1}\Delta t})+(-v_{1}+\frac{1}{v_{2}}+v_{2}) \ln\frac{r_{B}}{r_{A}}]
\simeq \nonumber \\ & & \frac{r_{g}E}{2}( \frac{1}{v_{2}}-\frac{1}{v_{1}}+v_{2}-v_{1})
\ln\frac{r_{B}}{r_{a}} {,}
\end{eqnarray}
where the approximation $v_{1}\Delta t\ll r_{A}$ has been used in deriving the last
result. On using the equation $1/v=E/p$ and the approximations $v\sim
1-\frac{m^{2}}{2E^{2}}-\frac{m^{4}}{8E^{4}}, 1/v\sim
1+\frac{m^{2}}{2E^{2}}+\frac{3m^{4}}{8E^{4}}$, one arrives at the final result
\begin{eqnarray} \label{2.62}
\lefteqn{(i\Delta \Phi_{g}\phi_{0})_{m_{1}}-(i\Delta
\Phi_{g}\phi_{0})_{m_{2}}\simeq}\nonumber
\\& &\frac{MGc^{5}}{4\hbar E^{3}}(m_{2}^{4}-m_{1}^{4}) \ln\frac{r_{B}}{r_{A}}= \nonumber
\\& &1.37\times 10^{-19}(\frac{M}{M_{\odot}})(\frac{\Delta m^{4}}{eV^{4}})(
\frac{MeV}{E}) ^{3}\ln\frac{r_{B}}{r_{A}} {.}
\end{eqnarray}
The effect therefore exists, but is extremely small in typical astrophysical
applications.

Torsion-induced neutrino oscillations have been considered by de Sabbata and
collaborators \cite{venzo3,venzo4}.-

\subsection{The equivalence principle and neutrino oscillations}

Gravitational fields can not generate neutrino oscillations if gravity couples
universally to matter. As first pointed out by Gasperini \cite{gasperini}, violations of
the equivalence principle could in principle affect the behaviour of neutrinos and be
tested in experiments on neutrino oscillations \cite{halprin}. Consider a spinless
particle in a Newtonian gravitational field. Its Hamiltonian in the WFA is
\begin{equation}\label{2.63}
H=(1-\gamma_{00}) ^{\frac{1}{2}}(p^{2}-\gamma^{ij}p_{i}p_{j}+m^{2})
^{\frac{1}{2}}-p_{i}\gamma^{i0} {,}
\end{equation}
which, in the simple case of a Newtonian potential, becomes
\begin{equation}\label{2.64}
H \sim p+\frac{m^{2}}{2p}-\frac{1}{2}p\gamma_{00} {,}
\end{equation}
where $\gamma_{00}=2\alpha\varphi(r) $ and $\alpha =1$ if the principle of equivalence is
not violated. Deviations from the equivalence principle are parameterized by assuming
that $\alpha\neq 1$ and takes different values for different neutrino mass eigenstates.
Assume the mass eigenstates $\nu_{1}, \nu_{2}$ are related to the weak eigenstates by the
transformation $ \left(\matrix{\nu_{1}\cr \nu_{2}}\right)= U \left(\matrix{\nu_{e}\cr
\nu_{\mu}}\right)$, where the unitary matrix
$U=\left(\matrix{\cos\theta&-\sin\theta\cr\sin\theta&\cos\theta}\right)$ represents
mixing in the two-generation case. The weak eigenstates then evolve according to the
equation
\begin{equation}\label{2.65}
i\frac{\partial}{\partial
t}\left(\matrix{\nu_{e}\cr\nu_{\mu}}\right)=U^{\dagger}(p+\frac{m^{2}}{2E}-\alpha
E)U\left(\matrix{\nu_{e}\cr\nu_{\mu}}\right) {,}
\end{equation}
where $ m^{2}=\left(\matrix{m_{1}^{2}&0\cr 0&m_{2}^{2}}\right)$, and
$\alpha=\left(\matrix{\alpha_{1}&0\cr 0&\alpha_{2}}\right)$. One therefore finds
\begin{equation}\label{2.66}
i\frac{\partial}{\partial t}\left(\matrix{\nu_{e}\cr
\nu_{\mu}\cr}\right)=M_{W}\left(\matrix{\nu_{e}\cr \nu_{\mu}\cr}\right) {,}
\end{equation}
\begin{eqnarray}\label{2.67}
M_{W}=\hspace{-.05in}\left(\matrix{\lefteqn{\hspace{-.8in}E+(
\frac{m_{1}^{2}}{2E}-E\varphi \alpha_{1}) \cos^{2}\theta\ + }\nonumber \\&
&\hspace{-4.0in}(\frac{m_{2}^{2}}{2E}-E\varphi\alpha_{2})
\sin^{2}\theta&\hspace{-1.9in}\vspace{.25in}(\frac{\Delta
m^{2}}{4E}-E\varphi\frac{\Delta\alpha}{2}) \sin(2\theta)\cr (\frac{\Delta m^{2}}{4E
}-E\varphi\ \frac{\Delta\alpha}{2}) \sin(2\theta)&
\lefteqn{\hspace{.3in}E+(\frac{m_{1}^{2}}{2E}-E\varphi\alpha_{1})
\sin^{2}\theta+}\nonumber
\\& & \hspace{.4in}(\frac{m_{2}^{2}}{2E}-E\varphi
\alpha_{2})\cos^{2}\theta\cr}\hspace{-.1in}\right){.}
\end{eqnarray}
Since the overall phase is unobservable, subtracting the constant
$E+(m_{1}^{2}-E\varphi\alpha_{1})\sin^{2}\theta+(m_{2}^{2}-E\varphi\alpha_{2})\cos^{2}\theta$
from the diagonal terms of $M_{W}$  does not affect oscillations in which only the
relative phases of the mass eigenstates are involved. This leads to the equation
\begin{equation}\label{2.68}
i\frac{\partial}{\partial
t}\left(\matrix{\nu_{e}\cr\nu_{\mu}}\right)=\frac{1}{2}(\frac{\Delta
m^{2}}{2E}-E\varphi\Delta\alpha)\left(\matrix{-2\cos(2\theta)&\sin(2\theta)
\cr\sin(2\theta)&0}\right)\left(\matrix{\nu_{e}\cr\nu_{\mu}}\right) {.}
\end{equation}
The solutions of (\ref{2.68}) are
\begin{eqnarray}\label{2.69}
\nu_{e}(t)&=& C_{1}e^{-i\omega t}+C_{2}e^{i\omega t}\nonumber \\
\nu_{\mu}(t)&=&D_{1}e^{-i\omega t}+D_{2}e^{i\omega t} {,}
\end{eqnarray}
with the condition $|\nu_{e}(t)|^{2}+|\nu_{\mu}(t)|^{2}=1$. One finds
$\omega=\frac{\Delta m^{2}}{4E}-E\varphi\frac{\Delta\alpha}{2}$. The initial condition
$\nu_{e}(0)=1$ is also used to determine the constants in (\ref{2.69}). One finds
$C_{1}=\sin^{2}\theta, C_{2}=\cos^{2}\theta, D_{1}=\sin\theta\cos\theta,
D_{2}=-\sin\vartheta\cos\theta$. The transition probability therefore is
\begin{equation}\label{2.70}
P_{\nu_{e}\rightarrow\nu_{\mu}}=\sin^{2}(2\theta)\sin^{2}(\omega t) {.}
\end{equation}
In the absence of gravity, $\alpha_{1}=\alpha_{2}=0$, flavour oscillations in vacuum
occur according to the MSW mechanism and are driven by $\Delta m^{2}$. The MSW
oscillations take place also when $\alpha_{1}=\alpha_{2}\neq 0$. On the other hand, when
gravity is present and $\alpha_{1}\neq\alpha_{2}\neq 0$, flavour oscillations occur not
only if $\Delta m^{2}\neq 0$, but also when $\Delta m^{2}=0$, with either $m_{1}=m_{2}$
or $m_{1}=m_{2}=0$ \cite{butler}. The charged-current interactions of $\nu_{e}$'s with
electrons in a star can also be taken into account by introducing the additional
potential energy $\surd 2G_{F}N_{e}(r)\leq 10^{-12}eV$. For the Sun
$N_{e}(r)=N_{0}exp(-10.54\frac{r}{R_{\odot}})cm^{-3}$, where $N_{0}$ is the number of
electrons at its centre \cite{bahcall}. Assuming $\Delta m^{2}=0$ for simplicity, the
equations of motion become in this case
\begin{eqnarray*}\label{2.71}
i\left(\frac{\partial}{\partial t}\right)\left(\matrix{\nu_{e}\cr\nu_{\mu}\cr}\right)=
\left(\matrix{\surd2G_{F}N_{e}(r)- \Delta\alpha
E\varphi\cos(2\theta)&\frac{\Delta\alpha}{2} E\varphi\sin(2\theta)
\cr\frac{\Delta\alpha}{2}
E\varphi\sin(2\theta)&0\cr}\right)\left(\matrix{\nu_{e}\cr\nu_{\mu}\cr}\right){.}
\end{eqnarray*}
The resonance condition $\surd2G_{F}N_{e}(r)=E\varphi\Delta\alpha\cos(2\theta)$
 is satisfied only when $\Delta\alpha<0$ because
 $\varphi<0$.

 \section{Summary}

 These lectures have dealt with non-relativistic and relativistic wave equations in weak,
 external gravitational and inertial fields. Only two fundamental aspects of the
 interaction have been considered: the generation of quantum phases and
 spin-gravity coupling.

 As shown in Section 2, quantum phases can be calculated exactly to first order in the
 field and in a manifestly covariant way for Klein-Gordon, Maxwell and Dirac equations.
 They can then be tested in experiments
 of increasing accuracy.

 The behaviour of quantum systems is consistent
 with that predicted by general relativity, intended as a theory of both gravity and
 inertia, down to distances $\sim 10^{-8}cm$. This is borne out of measurements on
 superconducting electrons $ (\sim 10^{-4}cm) $ and on neutrons $(\sim 10^{-8}cm) $ which are
 not tests of general relativity $per$
 $se$, but confirm that the behaviour of inertia and Newtonian gravity is that predicted by
 wave equations that satisfy the principle of general covariance. Atomic and molecular
 interferometers will push this limit down to $10^{-9}-10^{-11}cm $ and perhaps lead to new
 tests of general relativity. Prime candidates are in this regard a correction term to the
 gravitational red-shift of photons, a (general) relativistic correction to the
 gravitational field of Earth, and the Lense-Thirring effect of Earth. The last two
 experiments may require use of a near space laboratory to obviate the effect of rotation in
 the first instance and, on the contrary, to sense it in the latter case.

 The Mashhoon effect offers interesting insights into the interaction of inertia-gravity
 with spin. Spin is, of course, a quantum degree of freedom $par$ $ excellence$.  Spin-rotation
 coupling plays a
 fundamental role in precision measurements of the anomalous magnetic moment of muons.
 These extend the validity of the fully covariant Dirac equation down to distances
 comparable with the muon's wave-length, or $\sim 2\times 10^{-13}cm$.

 Rotational inertia does not couple universally to matter and does therefore
 violate the weak equivalence principle. This generates particle helicity oscillations
 that may play a role in some astrophysical processes.

 Other violations of the equivalence principle may play a role in neutrino oscillations.
 Even small violations, $\Delta\alpha \sim 10^{-14}$, could be amplified by a concomitant
 gravitational field. The resulting oscillations would then become comparable in
 magnitude with those due to the MSW effect.

 More serious violations of the equivalence
 principle would
 invalidate the principle of general covariance and render problematic the use of
 covariant wave equations.

\end{document}